\documentclass[a4paper,10pt,twocolumn,aps,pra,superscriptaddress,nofootinbib,longbibliography,showpacs]{revtex4-1}

\usepackage[utf8]{inputenc}
\usepackage[T1]{fontenc}

\usepackage{graphicx}
\graphicspath{{./fig/}{./tikz/}}

\usepackage{amssymb,amsmath}%,eulervm}

\newcommand*{\da}{\dagger}
\newcommand{\s}[1]{_\mathrm{#1}}

\newcommand*{\up}[1]{\ensuremath{^\text{#1}}}
\providecommand*{\Mode}[1]{\mathbf{#1}}
\usepackage[colorlinks,allcolors=blue]{hyperref}
\begin{document}

% >>>
% The title <<<
\title{Squeezer-based pulsed optomechanical interface}
\author{Andrey A. \surname{Rakhubovsky}}
\email{andrey.rakhubovskiy@upol.cz}
\author{Nikita \surname{Vostrosablin}}
\author{Radim \surname{Filip}}

\affiliation{Department of Optics, Palack{\'y} University, 17. Listopadu 12, 771 46 Olomouc, Czech Republic }

\begin{abstract}
	We prove feasibility of high-fidelity pulsed optomechanical interface based on all-optical pre-squeezing of non-Gaussian quantum states of light before they enter the optomechanical system. We demonstrate that feasible pre-squeezing of optical states effectively increases the low noise transfer of them to mechanical oscillator. It allows one to surpass the limit necessary to transfer highly nonclassical states with negative Wigner function. In particular, we verify that with this help single photon states of light can be efficiently turned to single phonon states of mechanical oscillator, keeping the negativity of the Wigner function. It opens the possibility to merge quantum optomechanics with the recent methods of quantum optics.
\end{abstract}

\date{\today}
\maketitle

% \listoftodos

% >>>

\section{Introduction} % <<<
\label{sec:introduction}

Recent development of continuous-variable tools of quantum optics~\cite{furusawa_quantum_2011} and quantum optomechanics~\cite{meystre_short_2013, chen_macroscopic_2013, aspelmeyer_cavity_2014} merges these two disciplines in one unique platform. Advantageously, both these fields can mutually benefit. A necessary step for their complete merger is the high fidelity transfer of nonclassical states of light to mechanical systems. Such quantum interface should be able to transfer a broad class of highly nonclassical states of light, for example, exhibiting negative Wigner function~\cite{lvovsky_quantum_2001,ourjoumtsev_quantum_2006,ourjoumtsev_generating_2006,neergaard-nielsen_generation_2006,neergaard-nielsen_high_2007,ourjoumtsev_generation_2007,bimbard_quantum-optical_2010,yukawa_generating_2013,yukawa_emulating_2013}. It is known that negative values of Wigner function are very fragile~\cite{caldeira_influence_1985,buzek_quantum_1995, zurek_decoherence_2003, deleglise_reconstruction_2008}; they can quickly disappear under influence of damping and noise, but also vanish in an inefficient interface.

Recently, entanglement between pulsed radiation and mechanical oscillator has been demonstrated~\cite{palomaki_entangling_2013}. It can be used to teleport quantum state of pulsed light in cavity optomechanical systems~\cite{hofer_quantum_2011}. However, teleportation strategy is not necessary for this purpose, since the state is not transferred at a distance. Instead of generating continuous-variable entanglement that is fragile under loss, we can directly use basic coupling between light and mechanical oscillator caused by a pressure of light.  Advantageously, this coupling provides a basic continuous variable gate --– quantum nondemolition (QND) interaction, when quantum states are strongly displaced before they start to interact with a mechanical system~\cite{aspelmeyer_cavity_2014}.

This non-demolition interaction in optomechanics has been already exploited in the regime of very short and intensive pulses to manipulate mechanical system without the cavity~\cite{vanner_pulsed_2011,machnes_pulsed_2012,vanner_cooling-by-measurement_2013,li_enhancing_2013}.  Very recently, the pulsed interface without the cavity based on multiple QND interactions has been proposed~\cite{bennett_quantum_2015}.
This platform exploits very short intensive pulses of light to reach sufficiently large optomechanical QND coupling existing beyond side-band resolved regime.
However, the QND interaction in the side-band resolve regime exhibits generally attractive potential of basic continuous variable gate with the nondemolition variables. From this reason, also the pulsed cavity optomechanics~\cite{hofer_quantum_2011} can advantageously use this type of interaction. Moreover, it can be simultaneously merged with cavity quantum optics, capable to produce and operate various non-Gaussian states of light~\cite{furusawa_quantum_2011}.

The strong and coherent displacement of a quantum state, achieved from long pumping pulses, is however not sufficient for its perfect upload as the pulse containing quantum state of light cannot be arbitrarily long. Consequently, the quantum interface is seriously limited by weak and slow coupling of light pulse to mechanical object.  Moreover, the interface suffers from residual thermal noise of mechanical oscillator, and additional technical noise and damping in the direct interface.  Although this interface does not necessarily break entanglement, it can be very limiting for a transmission of highly nonclassical quantum states of light. In general, quantum entanglement propagating through the interface can be enhanced by quantum distillation.
Quantum distillation of entanglement is only probabilistic, moreover, very demanding and practically requires quantum memories to improve the transfer of quantum states. For the continuous variable states, it moreover requires a venture beyond Gaussian operations and hence cannot be well applied here~\cite{eisert_distilling_2002,fiurasek_gaussian_2002,giedke_characterization_2002}.

It was already principally recognized that QND coupling between light and matter oscillator can be enhanced by a local pre-squeezing of quantum states of light before the coupling~\cite{filip_excess-noise-free_2008}. Interestingly, the mutual coupling to matter is enhanced purely by a \emph{local Gaussian} operation on light. In a fruitful combination with high-fidelity measurement of light and feedforward control of the oscillator, this allows one to achieve the optimal transfer of any, even non-Gaussian, state of light to matter oscillator. The transfer is suffering only from residual pure damping and all excess noise is in principle eliminated. Moreover, the residual damping can be made arbitrarily small as the pre-squeezing increases.

The squeezer-based QND interface can universally and deterministically transfer any state of light to mechanical oscillator. It is therefore different from conditional methods of preparation of non-Gaussian quantum states of mechanical systems~\cite{khalili_preparing_2010,mueller-ebhardt_quantum_2009,mueller-ebhardt_quantum-state_2012,akram_entangled_2013}. Furthermore, the proposed interface is capable of transferring of arbitrary states of light, without any prior knowledge about that state, which distinguishes it from recent proposals for preparation of mechanical oscillator in nonclassical states  (see~\cite{hammerer_nonclassical_2014} and references therein, and~\cite{paternostro_engineering_2011,sekatski_macroscopic_2014,galland_heralded_2014}). The ability to enhance the transfer by pre-squeezing makes the interface stand out from the ones relying on the beamsplitter-type optomechanical interaction~\cite{palomaki_coherent_2013, zhang_quantum-state_2003}.

Such the method can be extended to advanced QND scheme, which does not require the sophisticated cooling of the mechanical oscillator~\cite{marek_noise-resilient_2010}. These extensions are advantageous, because the procedures leading to better interface are fully deterministic and require only feasible Gaussian all-optical operations. Squeezing of single photon states and superposition of coherent states, both exhibiting negative Wigner function, have been already experimentally demonstrated~\cite{filip_measurement-induced_2005, miwa_exploring_2014}.

In this paper, we  investigate the application of this proof-of-principle approach to pulsed quantum optomechanics which is suitable for a merger with current optomechanical methods~\cite{hofer_quantum_2011} with already demonstrated online optical squeezer operating on the non-Gaussian states of light~\cite{miwa_exploring_2014}. We analyze the method beyond adiabatic elimination of the cavity mode and under the mechanical decoherence. We confirm that proof-of-principle idea can be applied to pulsed optomechanical systems. The squeezing is capable to obtain transmission of non-Gaussian states with negative Wigner function, when commonly used prolongation of coherent pulse is not helpful. We demonstrate this on a feasible example of squeezed single photon state transferred to the mechanical oscillator~\cite{meenehan_pulsed_2015}. This study certifies feasibility of merge of current quantum optics technology \cite{furusawa_quantum_2011} and developing quantum optomechanics.

% >>>
\section{Quantum non-demolition optomechanical coupling} % <<<
\label{sec:model_of_system}

\begin{figure}[t!]
	\centering
	\includegraphics[width=.65 \linewidth]{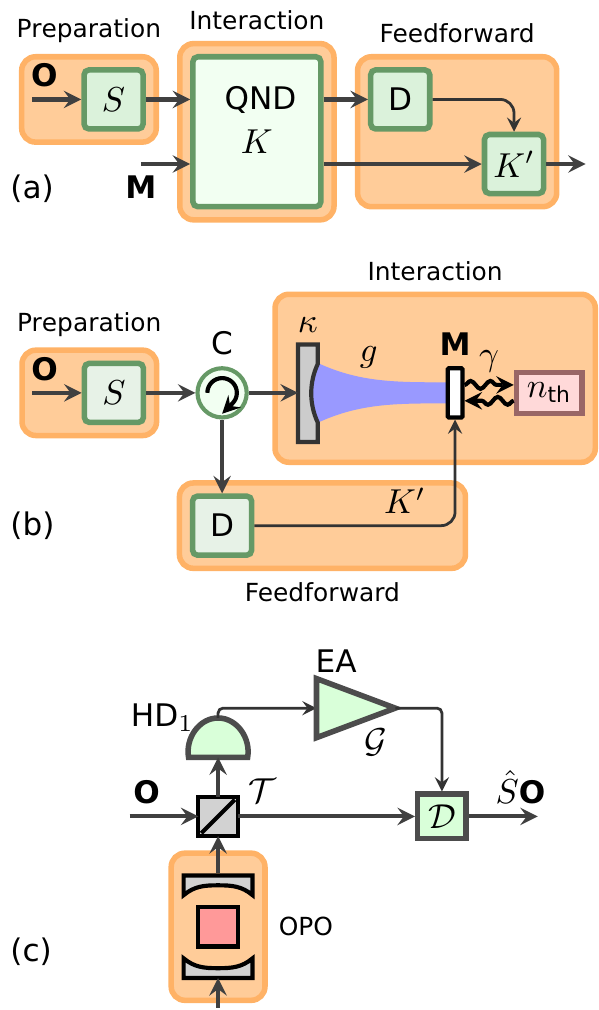}
	\caption{
		(a)~Simplified scheme of the interface. The optical mode (O) is prepared in the desired state, passes the squeezing operation ($S$) and is coupled via the QND interaction with the mechanical mode (M). The optical mode is then detected (D) and the outcome of the measurement is used to displace M.
		(b)~A sketch of optomechanical implementation, C --- circulator.
		(c)~Principal scheme of squeezing operation (see sec.~\ref{sec:pre_squezing_of_non_gaussian_states_of_light}), OPO --- optical parametric oscillator.
		% \\
		% Below: a sketch of optomechanical implementation.
	}
	\label{fig:protocol}
\end{figure}

We consider an interface allowing to transfer quantum state encoded in an optical pulse to the mechanical oscillator of an optomechanical system. The scheme of the interface is sketched at Fig.~\ref{fig:protocol} and mainly relies on the QND interaction in a cavity optomechanical system between the optical and mechanical modes comprising the system. The interaction is followed by detection on the optical side and consequent displacement of the mechanical mode based on the outcome of the detection.

The QND interaction with a macroscopic mechanical object was first proposed~\cite{caves_measurement_1980, braginsky_quantum_1980} to circumvent the standard quantum limit~\cite{braginsky_quantum_1995} of sensitivity of gravitational-wave detectors. This method was later revisited under the title of back-action evading measurement~\cite{clerk_back-action_2008} and has been recently realized experimentally~\cite{suh_mechanically_2014, lecocq_quantum_2015}.

The QND interaction between the two oscillators I and II is described by the Hamiltonian of the form $H\s{qnd} \propto g Q\s{I} Q\s{II}$, with $Q$ denoting \emph{quadrature} amplitudes of different oscillators, and $g$, coupling. In the rotating frame where the quadratures $Q$ are constants of motion the interaction of this type does not disturb them, but instead displaces the conjugate quadratures $P$ by an amount proportional to $g Q$. Consequently, prior squeezing of a mode that results in expansion of $Q$ is formally equivalent to an increase of the interaction strength. Using postsqueezing of the mode after the interaction, we can simply recover the QND with the increased interaction strength.

An optomechanical cavity can be thought of as a Fabry-P{\'e}rot resonator pumped through a semitransparent stationary mirror with another mirror being movable and perfectly reflective. The system thus comprises two harmonic oscillators (optical and mechanical modes) coupled via the radiation pressure. The Hamiltonian of the system reads~\cite{law_interaction_1995}:
\begin{equation}
	\notag
	H = \hbar \omega_c a^\da a + \hbar \omega_m b^\da b - \hbar g_0 a^\da a ( b^\da + b ),
\end{equation}
with $a$ ($b$) standing for annihilation operator of the optical (mechanical) mode with eigenfrequency $\omega_c$ ($\omega_m$).
The optomechanical coupling is inherently nonlinear and represents a nondemolition probe of the number of intracavity photons ($a^\da a$) by the displacement of the mechanics. This nondemolition nature of the interaction has been proposed for instance for detection of photon number~\cite{jacobs_quantum-nondemolition_1994} or increasing precision of thermal noise measurement~\cite{buchler_suppression_1999}.

In experimental realizations typically the single-photon coupling strength $g_0$ is small and the interaction is thus very weak. In order to observe it usually the system is considered in presence of a strong classical pump.  This allows to linearize the dynamics of the system and consider quantum fluctuations near the classical mean values. The Hamiltonian of the system is then written in the rotating frame defined by $H\s{rf} = \hbar \omega_c a^\da a + \hbar \omega_m b^\da b$ as follows:
\begin{equation}
	\notag
	H =  \hbar g_0 \sqrt{ n\s{cav}} ( a^\da e^{ - i \psi } + a e^{ i \psi }) ( b e^{ - i \omega_m t } + b^\da e^{ i \omega_m t } ),
	% \\
	% + H_p + H_d,
\end{equation}
where we assumed the pump to be resonant with the cavity. The phase $\psi$ is defined by the phase of the pump. The optomechanical coupling is enhanced by the mean optical amplitude proportional to the mean intracavity photon number $n\s{cav}$.

Assuming that the optical pump is modulated in such a way that $\sqrt{ n\s{cav} } \to \sqrt{ n\s{cav}} \cos ( \omega_m t + \phi )$, we apply rotating wave approximation (RWA) omitting terms rapidly oscillating at frequencies $\sim 2 \omega_m$ and obtain the following Hamiltonian
\begin{equation}
	\label{eq:onemorestep}
	H = \hbar g ( X \cos \psi - Y \sin \psi ) ( q \cos \phi - p \sin \phi ),
\end{equation}
where we defined the enhanced optomechanical coupling $g \equiv g_0 \sqrt{ n\s{cav}}$ and the optical and mechanical quadratures $X = ( a^\da + a ) $, $Y = i ( a^\da - a ) $, $q = ( b^\da + b ) $, and $p = i ( b^\da - b ) .$

A proper choice of phases $\phi$ and $\psi$ transforms~\eqref{eq:onemorestep} into a particular Hamiltonian corresponding to the QND interaction, for instance putting $\psi = 0,\  \phi = \pi$ yields $H_\text{I} = - \hbar g X q$.

Finally, to account for coupling to the environment we include viscous damping of the mechanical mode at rate $\gamma$, optical damping at rate $\kappa$, and noise terms~\cite{gardiner_quantum_2004} and write the Heisenberg-Langevin equations:
\begin{align}
	\label{eq:heizlang}
	\dot q & = - \frac{ \gamma }{ 2 } q + \sqrt{ \gamma } \xi _q,
	\\
	\notag
	\dot p & = - \frac{ \gamma }{ 2 } p + g X + \sqrt{ \gamma } \xi _p ,
	\\
	\notag
	\dot X & = - \kappa X + \sqrt{ 2 \kappa } X\up{in},
	\\
	\notag
	\dot Y & = - \kappa Y + g q +  \sqrt{ 2 \kappa } Y\up{in}.
\end{align}
Here $X\up{in}, Y\up{in}$ are the quadratures of the optical input mode, $\xi$ is the mechanical damping force with quadratures $\xi_{q,p}$ which we assume to be Markovian and satisfy usual commutation relations $[\xi_q (t) , \xi_p (t')] = 2 i \delta ( t - t')$.

Note that to write~\eqref{eq:onemorestep} we applied RWA at the mechanical frequency $\omega_m$ which requires the latter to exceed the rates of all the other processes taking place in the system, i.e., $\omega_m \gg \kappa , g , \gamma$. In practice it is sufficient to ensure the so-called \emph{resolved sideband regime}, $\omega_m \gg \kappa$. The experimental platform discussed here is therefore different from the one used in Ref.~\cite{bennett_quantum_2015}.

The system thus effectively comprises three modes: the input optical mode that encodes the target state, the intracavity optical mode, and the mechanical mode. The former two are coupled at rate $\kappa$ and the latter two are coupled at rate $g$. The intracavity mode thus serves as a transducer between the input optical and the mechanical modes. Under certain conditions the intracavity mode can be eliminated. In Sec.~\ref{sec:adiabatic_approximation} this elimination is performed to consider the system in a simple approximation. The system is considered without this elimination in Sec.~\ref{sec:full_solution} and the account of the mechanical bath is examined in Sec.~\ref{ssec:mechanical_decoherence_impact}.

% >>>
\section{Pre-squeezing of non-Gaussian states of light} % <<<
\label{sec:pre_squezing_of_non_gaussian_states_of_light}

During the past decade, quantum optics has progressed in the implementation of squeezing operation on quantum states of light. It was mainly due to development of the measurement-induced operations~\cite{filip_measurement-induced_2005}, which do squeeze any quantum state of light without injecting it into the cavity-based degenerate optical parametric amplifier. The basic scheme is depicted in Fig.~1(c). The input state of light is mixed with squeezed vacuum from OPO at the variable beam splitter with transmittivity~$\mathcal T$ and one output is measured by a high efficiency and low-noise homodyne detection HD$_1$. The electric signal from the detector is amplified in the electronic amplifier EA with the variable gain $\mathcal G$. It can be used to directly modulate ($\mathcal D$) the undetected output from the beam splitter in suitable optical quadrature. After optimization of $\mathcal G$ to eliminate noise of the non-squeezed variable from OPO, we can reach the transformation
\begin{equation}
	\notag
X\up{in}\rightarrow \frac{1}{\sqrt{\mathcal T}}X\up{in}, \quad
	Y\up{in}\rightarrow \sqrt{\mathcal T}Y\up{in}+\sqrt{1-\mathcal T}Y\up{sq}
\end{equation}
of the input operators $X\up{in},Y\up{in}$, where $Y\up{sq}$ is squeezed variable at the output of the OPO.
In the limit of sufficiently large squeezing produced by OPO, the input state can be intensively amplified in the variable $X\up{in}$ by the factor $S=1/\sqrt{\mathcal T}$, as has been demonstrated, for example, for the single photon state~\cite{miwa_exploring_2014}. The complementary variable $X\up{in}$ is  squeezed by the factor $S^{-1}$. Recently, dynamical control of the squeezing operation has been demonstrated~\cite{miyata_experimental_2014}. The purity of squeezed light from OPO is not limiting, because noise from antisqueezed quadrature can be eliminated in the feedforward loop. Recently, maximum squeezing from OPO reached $-12$ dB, which is sufficient to perform high-quality squeezing of non-Gaussian states of light. Moreover, recently fully optically integrated version of measurement-induced squeezer can improve phase stability and provide much higher quality of the squeezing procedure for very nonclassical states~\cite{masada_continuous-variable_2015}. Other improvements can be expected from the recent control of quantum states in optical cavities~\cite{yoshikawa_creation_2013}. The efficient schemes based on an optimal control of injection and extraction of non-Gaussian states in the cavity of OPO could in future substitute the measurement induced squeezers.

The feedforward strategy of all optical pre-squeezing can be further combined with the feedforward optomechanical interface. Instead to directly modulate light before it enters the optomechanical cavity, we can combine the results from homodyne measurement HD$_1$ with other results of homodyne detection D of light leaving the optomechanical cavity and apply them together to properly displace the mechanical state. The situation simplifies even more for a transfer of given state to mechanical oscillator. In this case, it is sufficient to prepare the squeezed version of this state directly, for example, using recent high-fidelity tunable multi-photon subtraction schemes. The squeezing on the top of non-Gaussian states can be very large, up to already experimentally generated $-12$ dB~\cite{mehmet_squeezed_2011}. Although this method is conditional and not universal, it can be versatile for high-quality preparation of non-Gaussian quantum state of mechanical system. The simplest testing situation appears if the highly squeezed state is transferred to the mechanical oscillator. The squeezing is then used to prepare a ground state or squeezed state of mechanical system. It can be done differently, using the projection by homodyne measurement D, than recently demonstrated squeezing in electromechanical oscillators~\cite{wollman_quantum_2015, pirkkalainen_squeezing_2015}.

% >>>
\section{Adiabatic elimination of intra-cavity mode} % <<<
\label{sec:adiabatic_approximation}

Before we present full analysis, we repeat the basic idea of the squeezer-based interface~\cite{filip_excess-noise-free_2008} in the simplest approximation, where the cavity mode is adiabatically eliminated and mechanical bath is not very occupied. It allows us to simply imagine the ideal performance of the squeezer-based optomechanical interface.

Typically in an optomechanical experiment $\kappa \gg  \gamma/2$ holds, so if the mechanical bath is not very occupied, one could assume $\gamma = 0$. The condition $\kappa \gg g$ is commonly satisfied as well, which means that the optical mode in the cavity can respond to any changes in the input mode or the mechanical mode instantaneously [this amounts to putting $\dot X = \dot Y = 0$ in~\eqref{eq:heizlang}].

Consequently the intracavity mode is removed and in this simple picture the QND interaction between the input optical and mechanical modes results in exchange of one of the quadratures
\begin{equation}
\begin{aligned}
	\label{eq:aetr}
	q (\tau) & = q(0),
	&
	p (\tau) & = p(0) + K \Mode X\up{in},
	\\
	\Mode X\up{out} & = \Mode X\up{in},
	&
	\Mode Y\up{out} & = \Mode Y\up{in} + K q (0),
\end{aligned}
\end{equation}
with transfer coefficient
\begin{equation}
	\label{eq:Kdef}
	K = g \sqrt{ 2 \tau / \kappa }
\end{equation}
To write the transformations~\eqref{eq:aetr} we defined the quadratures of input and output pulses as integral over the rectangular pulse
\begin{equation}
	% \label{eq:iodef}
	\notag
	\Mode Q\up{k} = \frac{ 1 }{ \sqrt \tau } \int_0^\tau ds \: Q\up{k} (s),
	\quad Q = X , Y,
	\quad \text{k} = \text{in,out},
\end{equation}
so that $[\Mode X\up{k}, \Mode Y\up{k}] =  2 i$, and used input-output relations $Q\up{out} = \sqrt{ 2 \kappa } Q - Q\up{in}$.

To complete the state transfer to the mechanical mode we need to upload $\Mode Y\up{in}$ to the mechanical quadrature $q$. This can be achieved by a feedforward displacing the mechanical mode by amount equal to $ - K' \Mode Y\up{out}$. The feedforward control of mechanical oscillator in an optomechanical cavity was realized in several setups, for instance, by means of radiation pressure or dielectric gradient force actuation~\cite{cohadon_cooling_1999, arcizet_radiation-pressure_2006, kleckner_sub-kelvin_2006, poggio_feedback_2007, harris_feedback-enhanced_2012}.

The feedforward can in principle be implemented by a QND interaction with the second pulse via the Hamiltonian $H_\text{II} = - \hbar g' Y p$. If the duration of the second pulse equals $\tau'$, the coupling will be analogous to~\eqref{eq:aetr} with transfer coefficient $K' = g' \sqrt{ 2 \tau' / \kappa }$. After this procedure the two quadratures of the input optical mode are written to the quadratures of the mechanical mode
\begin{equation}
	\notag
\begin{aligned}
	q' & = q(0) ( 1 - K K' ) - K' \Mode Y\up{in},
	\\
	p' & = p(0) + K \Mode X\up{in}.
\end{aligned}
\end{equation}

The quadratures of the optical pulse are transferred to the mechanical mode in two steps, one quadrature at time. Therefore, squeezing of the pulse that amplifies one of the quadratures at cost of reduction of the other one can help to transfer the amplified quadrature. Indeed, squeezing of the pulse amounts to substitution $\Mode X\up{in} \to S \Mode X\up{in};\ \Mode Y\up{in} \to S^{-1} \Mode Y\up{in}$ in~\eqref{eq:aetr} and this is equivalent to replacement $K \to S K$.

The other quadrature transfer can be enhanced by increasing the feedforward gain. Or, if we think of the feedforward as of another QND interaction with limited strength, by appropriate amplification of the optical mode. Note, that the squeezing of the first pulse weakens the quadrature $\Mode Y\up{in}$ that should be transfered in the second step, so the amplification should account for it, which means, the gain of the feedforward should be replaced as $K' \to K' S$. However, since the other quadrature $\Mode X\up{out}$ is no longer of our interest, the amplification needs not to be noiseless as long as the noises are concentrated in this other quadrature.

After the feedforward the mechanical mode contains the squeezed target state:
\begin{equation}
	\label{eq:adiabatic_io}
	\begin{aligned}
		q_f = \sqrt \frac{ K' }{ K S } \left[ q (0) \sqrt{ 1 - T } - \sqrt{ T } \Mode Y\up{in} \right];
		\\
		p_f = \sqrt \frac{ K S }{ K' } \left[ p (0) \sqrt{ 1 - T } + \sqrt{ T } \Mode X\up{in} \right],
	\end{aligned}
\end{equation}
with transmittivity
\begin{equation}
	% \label{eq:Tdef}
	\notag
	T =  \frac{ (K S )^2 }{ 1 + ( K S )^2 },
\end{equation}
provided that $ K' = \frac{ K S }{ 1 + ( K S )^2 }$.

The transfer coefficient thus depends only on the product $K S$ and increasing this product allows to approach an ideal transfer with $T = 1$.  From the definition~\eqref{eq:Kdef} of $K$ it follows that for a given $\kappa$ the same increase in the product $K S$ can be provided by means of equal increase of either $S$, $g$ or $\sqrt\tau$. Increasing coupling strength or duration of the pulse can impose difficulties in experimental optomechanical realization. At the same time stronger presqueezing of the optical pulse helps to improve the transfer by cost of additional resource of external quantum optical tools.

The equations~\eqref{eq:adiabatic_io} can be recast in terms of target state and added noise (we \emph{formally} consider squeezing of the mechanical state to symmetrize the expressions)
\begin{equation}
	\label{eq:beamsplitter}
	q_f = - \sqrt{ T } \Mode Y\up{in} + \sqrt{ 1 - T } X_N,
	p_f = \sqrt{ T } \Mode X\up{in} + \sqrt{ 1 - T } Y_N
\end{equation}
The transformation is effectively combining the target state with quadratures $\Mode X\up{in}, \Mode Y\up{in}$ and a noisy mode with quadratures $X_N,Y_N$ on a beam splitter with the transmittivity $T$.

The variance of the added noise is defined as the product of the variances of quadratures of the noisy mode
\begin{equation}
	\label{eq:vndef}
	V_N \equiv \sqrt{ V_{X_N} V_{Y_N} }
\end{equation}
and is limited (see~\cite{caves_quantum_1982}) from below by the shot noise level: $V_N \geqslant 1$. Protocol that saturates the inequality is said to realize the \emph{excess-noise-free} upload~\cite{filip_excess-noise-free_2008}. This excess-noise-free is very advantageous for transmission of nonclassical features of non-Gaussian states, like transfer of single photon states to single phonon state. It is due to much higher robustness of quantum non-Gaussianity to loss than to the phase-insensitive noise.

The transfer defined by~\eqref{eq:adiabatic_io} represents mixture of the target state with the initial state of the mechanical mode and is excess-noise-free, provided the mechanical mode is initially in the ground state.

% >>>
\section{Beyond adiabatic approximation of optomechanical coupling} % <<<
\label{sec:full_solution}

In the previous section we demonstrated the principal possibility of an excess-noise-free transfer of a quantum state of light to mechanics that can be enhanced by optical presqueezing. The necessary condition of the transfer was the instantaneous reaction of the cavity mode to any changes. In this section we perform analysis showing that the cavity memory effects caused by the finite cavity reaction time do not limit the interface performance.  We first analyze the system assuming no mechanical decoherence ($\gamma = 0$) and then carry out the full analysis in Sec.~\ref{ssec:mechanical_decoherence_impact}.

After interacting with a presqueezed pulse, the mechanical system has quadratures
\begin{align}
	\notag
	q (\tau) & = q(0),
	\\
	\notag
	p (\tau) & = p (0) 	+
	K S \Mode X\up{in}
	\\
	\notag
	& + \frac g \kappa (1 - e^{ - \kappa \tau }) X (0)
	- \frac{ g S }{ \kappa } \sqrt{ 1 -e^{ - 2 \kappa \tau }} \Mode X\up{in}_\delta.
\end{align}

Due to the cavity memory effect, the simple transformations~\eqref{eq:aetr} become disturbed by the quadrature of the intracavity mode $X(0)$ and an auxiliary asymmetric mode of input field $\Mode X\up{in}_\delta$, defined as
\begin{equation}
	\label{eq:deltadef}
	\Mode Q\up{in}_\delta = \sqrt{ \frac{ 2 \kappa}{ 1 - e^{ - 2 \kappa \tau }}} \int_0^\tau ds \: e^{ - \kappa ( \tau - s ) } Q\up{in} (s ), \ Q = X, Y,
\end{equation}
in order to satisfy commutations $[ \Mode X\up{in}_\delta, \Mode Y\up{in}_\delta ] =  2 i$.

By definition the asymmetric mode is composed primarily of the values of $Q\up{in} (t)$ adjacent to the end of the interval of integration, which is a manifestation of memory. In the limit where the cavity mode could be eliminated, $ \kappa \to + \infty$ the integration kernel approaches Dirac delta [$e^{ - \kappa ( \tau - s )} \sim \delta ( \tau - s )$], so $\Mode X\up{in}_\delta \sim X\up{in} (\tau)$ up to normalization. Finally, the prefactor makes contribution of this term negligible.

The input-output transformation for the optical quadrature $\Mode Y\up{out}$ that is of our interest reads
\begin{multline}
	\notag
	\Mode Y\up{out} = q (0) K \left( 1 - \frac{ 1 - e^{ - \kappa \tau }}{ \kappa \tau } \right)
	+ S^{-1} \Mode Y\up{in}
	\\
	+ Y (0) \sqrt{ \frac{ 2 }{ \kappa \tau } } (1 - e^{ - \kappa \tau } ) - S^{-1}  \sqrt{ \frac{ 2 }{ \kappa \tau } ( 1 - e^{ - 2 \kappa \tau }) } \Mode Y\up{in}_\delta.
\end{multline}
We would like to note that the cavity memory effect manifests itself differently in the optical output and the mechanical modes, which makes introduction of the asymmetric mode in the form~\eqref{eq:deltadef} necessary.

After the displacement $q (\tau) \to q(\tau) - K' S \Mode Y\up{out}$ and \emph{formal} postsqueezing with amplitude $\sqrt{ K S / K' }$, the mechanical mode has quadratures
\begin{align}
	\notag
	q_f & = - \sqrt{ T } \Mode Y\up{in} + \sqrt{ 1 - T }
	\bigg[
	q (0) \left( 1 + \frac{ 2 g^2 S^2 }{ \kappa^2 } ( 1 - e^{ - \kappa \tau } )  \right)
	\\
	\notag
	& - Y(0) \frac{ 2 g S^2 }{ \kappa } ( 1 - e^{ - \kappa \tau })
	+ \Mode Y\up{in}_\delta \frac{ 2 g S }{ \kappa } \sqrt{ 1 - e^{ - 2 \kappa \tau}}
	\bigg];
	\\
	\notag
	p_f & = \sqrt T \Mode X\up{in} + \sqrt{ 1 - T }
	\bigg[
	p (0)
	+ X (0) \frac{ g }{ \kappa } ( 1 - e^{ - \kappa \tau } )
	\\
	& - \Mode X\up{in}_\delta  \frac{ g S }{ \kappa } \sqrt{ 1 - e^{ - 2 \kappa \tau }}
	\bigg].
	\label{eq:mech_final}
\end{align}
The presqueezing of the incoming pulse is thus not completely equivalent to increasing coupling strength due to the fact that the squeezing changes the ratio in $\Mode Y\up{out}$ of the uploaded and intracavity modes in favor of the latter.

\begin{figure}[t]
	\centering
	\includegraphics[width = .95 \linewidth]{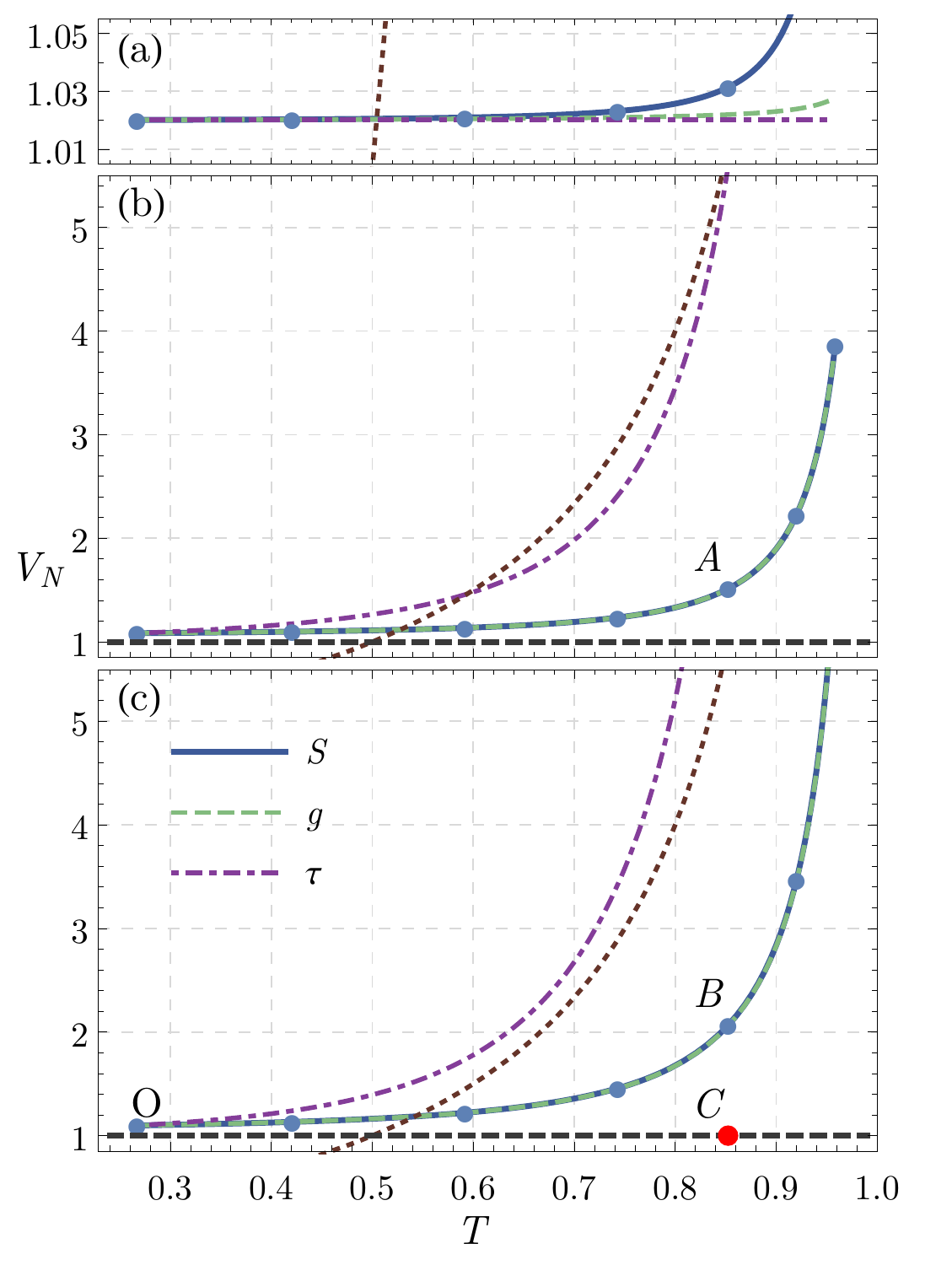}
	\caption{
		Added noise variance $V_N$ (\ref{eq:beamsplitter},\ref{eq:vndef}) in shot noise units as a function of the transfer coefficient $T$ in case of increase of squeezing $S$, coupling $g$, or pulse duration $\tau$.
		(a) Solution without the mechanical bath (Sec.~\ref{sec:adiabatic_approximation}); (b) solution accounting for the mechanical bath but with adiabatically eliminated cavity mode; (c) full solution (Sec.~\ref{ssec:mechanical_decoherence_impact}).
		The point (O) corresponds to the initial set of parameters~\eqref{eq:parameters} with no presqueezing ($S=1$). Markers denote points corresponding to sequential increase of squeezing with the step $3$ dB.  Dashed line marks the perfect excess-noise-free transfer $V_N=1$. Dotted curve indicates the maximum $V_N$ allowing transfer of the negativity of the Wigner function of a single-photon state for a given transmittivity $T$ [$V_N^2 < T/ (1 - T)$].
	}
	\label{fig:vn_full}
\end{figure}

The added noises in this scheme are provided by the initial occupation of the optical and mechanical modes of the optomechanical cavity, and the asymmetric mode caused by the finite cavity decay $\kappa$. Assuming all the noise modes in ground state, the added noise variance $V_N$ can be approximated by
\begin{equation}
	\notag
	V_N \approx \sqrt{ 1 + 4 g^2 S^4 / \kappa^2 }.
	% Seriously, the variance equals square root of this; a bit more exact
	% solution is
	% V_N^2 \approx 1 + ( 9 + 4 S^2 ) g^2 S^2 / \kappa^2 .
\end{equation}
Therefore, for high quality low-noise transfer we need to secure  $g S^2 / \kappa \ll 1$ that keeps the added noise close to the vacuum level and  $g S \sqrt{ \tau / \kappa } \gg 1$ for high transmittivity. Both inequalities are satisfied by making pulses longer: $\tau \gg 1/ g$.  This is illustrated at Fig.~\ref{fig:vn_full}(a) where we plot the added noise variance as a function of the transmittivity of the interface for the cases of increasing squeezing, coupling or pulse duration. It is evident from the figure and from~\eqref{eq:mech_final} that increasing $\tau$ allows one to increase transfer while adding little excess noise. At the same time increasing coupling or squeezing adds more noise than increasing $\tau$. It is due to the intracavity mode which disturbs ideal dynamics observed in the adiabatic approximation. The larger $\tau$ is, however, prolonging the interaction which requires better phase stability of the transfer and, mainly,
smaller decoherence potentially caused by mechanical environment of the oscillator.
To make a proper conclusion on the choice of the best strategy, therefore, we need to take into account the mechanical bath.

% >>>
\section{Transfer under mechanical decoherence} % <<<
\label{ssec:mechanical_decoherence_impact}

In this section we consider an imperfect QND interaction of a pulse with the system during which the mechanical mode is affected by its bath. Simultaneously, we keep the analysis beyond the adiabatic approximation. We still consider that highly efficient homodyne measurement is followed by a perfect instantaneous electro-mechanical feedforward. We first present an approximate analysis to get order-of-magnitude estimates, then sketch the steps to obtain full analytical solution, and provide its results.

In the case of small mechanical damping $\gamma \ll g, \kappa$ the simplest estimates can be easily obtained from Eqs.~\eqref{eq:mech_final} by the substitution
\begin{equation}
	\notag
	Q(0) \to Q(0) + \sqrt{ 1 - e^{ - \gamma \tau }} Q_B,
	\quad
	Q = q , p,
\end{equation}
where $Q_B$ represents effective quadratures of the mechanical bath with variance $2 n\s{th} + 1$, with $n\s{th}$ being the average bath occupation. The corresponding contribution of the thermal noise to the variances of each of the added noise terms $X_N, Y_N$ is  $ \gamma^2 \tau^2 ( 2 n\s{th} + 1 ) / ( 1 - T )$.

In the region of parameters where the thermal noise from the bath dominates making other noises negligible, the added noise variance can be approximated by $V_N = 1 + 2 \gamma ( g \tau S )^2 ( 2 n\s{th} + 1 ) / \kappa $. Despite the fact that $g$, $\tau$ and $S$ enter this expression equally, we note that $T \propto g S \sqrt \tau$, and hence increase in pulse length that produces the same increase in transmittivity, adds much more thermal noise then stronger coupling or squeezing.

We therefore can write several asymptotic requirements for high-transmittivity low-noise state transfer. First, to achieve high transfer gain, we need $g^2 S^2 \tau / \kappa \equiv \epsilon \gg 1$. Second, in order to make cavity mode induced effects negligible, $g^2 S^4 / \kappa^2 = S^2 \epsilon / ( \kappa \tau ) \ll 1$. Finally, to keep the thermal noise influence moderate, $\gamma ( g \tau S )^2 n\s{th} / \kappa = \gamma \tau n\s{th} \epsilon \ll 1$. The two latter combine to set the proper range for the available values of $\tau$:
\begin{equation}
	\notag
	\frac{ S^2 \epsilon }{ \kappa } \ll \tau \ll \frac{ 1 }{ \epsilon \gamma n\s{th} }.
\end{equation}

% >>>
% \subsection{Accurate solution} % <<<
% \label{ssec:accurate_solution}

Along with the simplest estimates one can perform a full analysis of the system dynamics to properly quantify the impact of different sources of the noise.

The system of dynamical equations~\eqref{eq:heizlang} is linear and therefore has~\cite{korn_mathematical_2000} a formal analytical solution that involves exponential of the matrix of its coefficients. The solution is rather complicated, and so we will present here only the expression for $p(\tau)$
\begin{align}
	\notag
	p (\tau) & = p(0) e^{ - \gamma \tau /2 } + \int_0^\tau ds e^{ \gamma ( s - \tau) / 2 } \sqrt{ \gamma } \xi_p (s)
	% \\
	%&
	+ \theta ( \tau ) X(0)
	\\
	\notag
	& + K S \Mode X\up{in}
	- \int_0^\tau ds \left[ \frac{ K }{ \sqrt{ \tau }} - \sqrt{ 2 \kappa } \theta ( \tau - s ) \right] S X\up{in} (s ),
	\\
	\notag
	\theta (t) & \equiv \frac{ g }{ \kappa - \frac{ \gamma }{ 2 } }  \left( e^{ - \gamma t / 2 }  -e^{ - \kappa t } \right).
\end{align}
The last summand in the expression for $p (\tau)$ represents the asymmetric mode modified by mechanical decoherence. In the limit $\gamma \to 0$ this summand is reduced to $\Mode X\up{in}_\delta$.

Using the formal solution one can write expressions for $q(\tau)$ and $\Mode Y\up{out}$ and proceed further to obtain the beamsplitterlike transformations in the form~\eqref{eq:beamsplitter}. With this transformation one can analyze the added noise variance $V_N$~\eqref{eq:vndef}.

Our result of estimation for the added noise variance is presented in Fig.~\ref{fig:vn_full}(c). For estimations we used the following initial set of parameters:
\begin{gather}
	\label{eq:parameters}
	\kappa = 221.5 \text{ MHz},\ g = 1 \text{ MHz},\ \gamma = 328 \text{ Hz},
	\\
	\notag
	\tau = 4 \times 10^{-5} \text{ s}
\end{gather}
of a recent reported experiment with optomechanical crystal~\cite{meenehan_pulsed_2015}. The mechanical system is very well precooled in this experiment; however, there is intensive heating of the mechanical mode by optical pump already at the level of circulating power equivalent to a few photons. We model this by setting initial mechanical occupation to $n_0 = 0.01$ and mechanical bath occupation to $n\s{th} = 2$.

Starting from the set~\eqref{eq:parameters} that is represented by the point $O$ in the figure, we continuously increased one of the parameters $S, g$ or $\tau$. Comparing Fig.~\ref{fig:vn_full}(c) with Fig.~\ref{fig:vn_full}(a) allows one to conclude that for the parameters we chose the main source of added noise is indeed the thermal mechanical environment. In this case, optical presqueezing is effectively equivalent to increasing the interaction strength as it follows from the simple estimate. This is as well seen from Fig.~\ref{fig:vn_full}(c), where the curves corresponding to increase in $S$ and $g$ overlap.

The effect of the intracavity optical mode has two contributions. First, this mode itself produces some excess noise and, second, it serves as a memory that enhances the impact of the thermal noise. To illustrate this we analyze the transfer under the mechanical decoherence but adiabatically eliminating the cavity mode. The result of this analysis is presented at Figs.~\ref{fig:vn_full}(b). Although the mechanical bath is the main source of noise in both Fig.~\ref{fig:vn_full}(b) and~(c), it is clearly seen that elimination of the cavity mode leads to underestimation of the noise impact.

\begin{figure}[t]
	\centering
	\includegraphics[width=.95\linewidth]{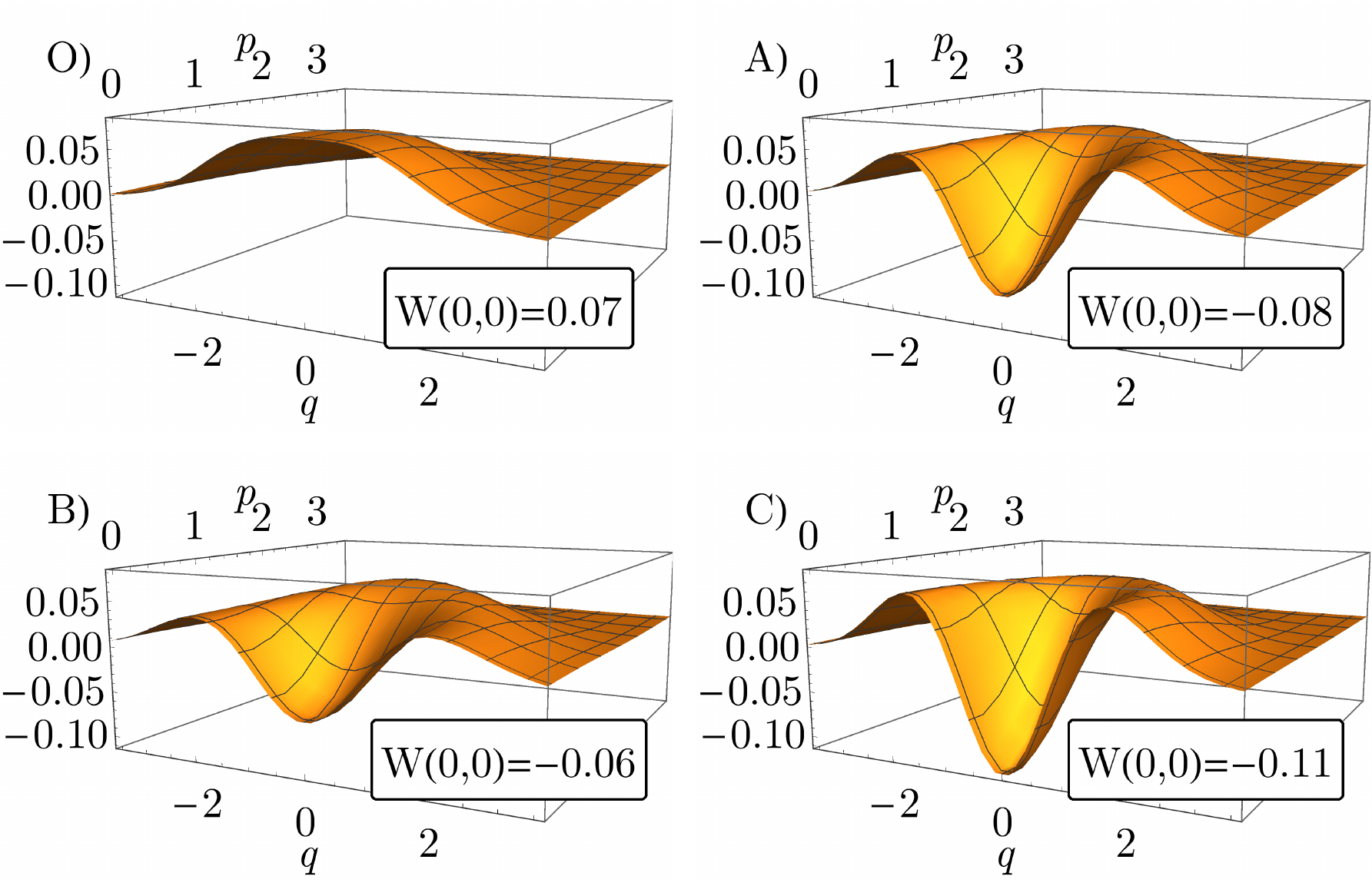}
	\caption{Wigner function of a Fock state $|1 \rangle$ transmitted to the mechanical system using the proposed protocol working in the regimes, denoted by corresponding named points at Fig.~\ref{fig:vn_full}. $O$, without the presqueezing; $A$ with $12$ dB of presqueezing with intracavity mode formally eliminated; $B$ with $12$ dB of presqueezing, full solution; $C$, same transmittivity as in $A$ and $B$, but no excess noise ($V_N = 1$).
	The negativity at the origin is presented at the framed insets.
	}
	\label{fig:wigner}
\end{figure}

The dotted lines in Fig.~\ref{fig:vn_full} denote the maximum variance of added noise for a given transmittivity that allows one to transfer negativity of a single-photon state. Our analysis shows that the interface we consider is capable of such a transfer. It is due to possitive effect of the pre-squeezing, which allows one to shorten the interaction time. From this reason, the squeezer-based pulsed optomechanical interfaces in the cavities is a feasible road to achieve high-fidelity transfer of the non-Gaussian quantum states of light to mechanical oscillators. Indeed, the Wigner function of an uploaded state manifests negativity at the origin. The simplest example is transfer of a highly nonclassical single photon state to a single phonon state.
This is demonstrated in Fig.~\ref{fig:wigner}, where we compare the Wigner functions of a single photon state, transferred with help of the proposed protocol in regimes, denoted in Fig.~\ref{fig:vn_full} by the points $O$, $A$, $B$ with the state uploaded via the excess-noise free protocol ($V_N = 1$, point $C$). We clearly observe a detectable preservation of negativity of the Wigner function of a single phonon state. It witnesses high-fidelity quantum transfer preserving effects which cannot be explained by stochastic mechanics. Based on this example, we can conclude that squeezed-based pulsed optomechanical interface is a feasible road to single photon-phonon transfer. Moreover, it can be used more generally to transfer other non-Gaussian states of light to mechanical systems.

We did not consider mechanical decoherence during the feedforward control of the mechanical system. This simplification is not very coarse, even if we consider a second QND interaction in place of the feedforward. In our protocol interaction gains $K$ and $K'$ relate to each other as $K' \propto K^{-1}$. Therefore with increase $K \gg 1$ which is a natural condition to achieve $T$ close to $1$, the gain of the second interaction $K'$ and consequently its duration $\tau'$ decrease. The second interaction therefore effectively approaches an instantaneous feedforward that does not suffer much from the thermal noise.

% >>>
\section{Conclusion} \label{sec:conclusion} % <<<

We have verified feasibility and performance of the squeezer-based high-fidelity optomechanical interface for deterministic transfer of non-Gaussian highly nonclassical quantum states of light to mechanical oscillators. We observed clearly that interfaces which cannot transfer negativity of Wigner function can be improved by this method to be able to preserve it. We demonstrated importance of verification beyond the adiabatic elimination. We proved that squeezer-based interface is especially useful when the transfer is influenced by mechanical decoherence, which limits time duration of transfer. We predicted achievable quality of the interface for the experiment~\cite{meenehan_pulsed_2015}. This interface merges developing pulsed cavity optomechanics~\cite{aspelmeyer_cavity_2014} with recent state-of-the-art of continuous-variable quantum optics~\cite{furusawa_quantum_2011}. It opens therefore a new joint direction of cavity-based quantum optomechanics and cavity-based quantum optics. In this joint direction, the fields can be mutually fruitful and produce a united physical platform for new continuous-variable experiments with nonclassical light and mechanical oscillators.

% >>>
% Acknowledgments %<<<
\begin{acknowledgments}
We acknowledge Project No. GB14-36681G of the Czech Science Foundation.
The research leading to these results has also received funding
from the EU FP7 under Grant Agreement No. 308803
(Project BRISQ2), co-financed by M\v SMT \v CR (7E13032). A.A.R. acknowledges support by the Development Project of Faculty of Science, Palacky University. N.V. acknowledges the support of Palacky University (IGA-PrF-2015-005).
\end{acknowledgments}%>>>

\bibliography{optomech_filtered}

\end{document}